\documentclass[12pt,a4,authoryear]{article}
\usepackage[authoryear]{natbib}
\usepackage{authblk}
\usepackage{amsmath}
\usepackage{amsthm}
\usepackage{xcolor}
\usepackage{CJK}
\usepackage{blkarray}
\usepackage{geometry}
\usepackage{mathrsfs}
\usepackage{booktabs}
\usepackage{epstopdf}
\usepackage{diagbox}
\usepackage{amsfonts}
\usepackage[colorlinks,
linkcolor=blue,
anchorcolor=blue,
citecolor=blue
]{hyperref}
\usepackage{multicol}
\usepackage{multirow}
\usepackage{algorithm}
\usepackage{algorithmic}
\usepackage{rotating}
\usepackage{setspace}
\usepackage{graphicx}
\usepackage[toc,page,title,titletoc,header]{appendix}
\usepackage{indentfirst} 
\newtheorem{thm}{Theorem}

\allowdisplaybreaks[4]

\def\squarebox#1{\hbox to #1{\hfill\vbox to #1{\vfill}}}
\def\boxit#1{\vbox{\hrule\hbox{\vrule\kern6pt
          \vbox{\kern6pt#1\kern6pt}\kern6pt\vrule}\hrule}}

\begin{document}
	\title{Reconstruct Kaplan--Meier Estimator as M-estimator and Its Confidence Band}
	\author[1]{Jiaqi Gu}
	\author[1]{Yiwei Fan}
	\author[1]{Guosheng Yin}
	\affil[1]{Department of Statistics and Actuarial Science, The University of Hong Kong
	}
	\date{}                     
	\setcounter{Maxaffil}{0}
	\renewcommand\Affilfont{\itshape\small}
	\maketitle
	
	\begin{abstract}
		The Kaplan--Meier (KM) estimator, which provides a nonparametric estimate
of a survival function for time-to-event data, has wide application in clinical studies, engineering, economics and other fields.
The theoretical properties of the KM estimator including its consistency and asymptotic distribution have been 
extensively
studied.
We reconstruct the KM estimator as an M-estimator 
by maximizing a quadratic M-function based on concordance,
which can be computed using the expectation--maximization (EM) algorithm. 
It is shown that the convergent point of the EM algorithm coincides with the traditional KM estimator, 
offering a new interpretation of the KM estimator as an M-estimator. 
Theoretical properties including the large-sample variance and limiting distribution of the KM estimator 
are established using M-estimation theory. Simulations and application on two real datasets 
demonstrate that the proposed M-estimator is exactly equivalent to the KM estimator, while 
the confidence interval and band can be derived as well.

	\end{abstract}

\noindent
	
	\section{Introduction}
In the field of clinical studies, analysis of time-to-event data is of great interest \citep{altman1998time}. 
The time-to-event data record the time of an individual from entry into a study till the occurrence of an event of interest,
such as the onset of illness, disease progression, or death.
In the past several decades, various methods have been developed for time-to-event data analysis, including the Kaplan--Meier  
(KM) estimator \citep{kaplan1958nonparametric}, the log-rank test \citep{Mantel1966} and the Cox proportional
hazards model \citep{Cox1972,Breslow1974}. Among these methods, the KM estimator is 
the most widely used nonparametric method to estimate the survival curve for time-to-event data. As a step function with jumps at the time points of observed events, the KM estimator is very useful to study the survival function of the event of interest (e.g. disease progression or death) when loss to the follow-up exists. By comparing the KM estimators of treatment and control groups, patients' response to treatment over time can be compared. Other than public health, medicine and epidemiology,
the KM estimator also has broad application in other fields, 
including engineering \citep{huh2011adaptive}, economics \citep{danacica2010using} and sociology \citep{kaminski2012survival}.
	
The KM estimator is well developed as a nonparametric maximum likelihood estimator \citep{johansen1978product}. As a result, asymptotic theories of the KM estimator have been extensively discussed in the literature. \citet{greenwood1926report} derived Greenwood's formula for the large-sample variance of the KM estimator at different time points and the consistency of the KM estimator is shown by \citet{peterson1977expressing}. By estimating the cumulative hazard function with the Nelson--Aalen estimator, \citet{Breslow1974} proposed the Breslow estimator which is asymptotically equivalent to the KM estimator.
The KM estimator converges in law to a zero-mean Gaussian process whose variance-covariance function
can be estimated using Greenwood's formula. In Bayesian paradigm, \citet{Susarla1976} proved that the KM estimator is a limit of the Bayes estimator under a squared-error loss function when the parameter of the Dirichlet process prior $\alpha(\cdot)$ satisfies $\alpha(R^+)\to 0$.
	
	In this paper, we develop an M-estimator for the survival function which can be obtained recursively via the expectation--maximization (EM) algorithm. When the M-function is quadratic, we show that the traditional KM estimator is the limiting point of the EM algorithm. As a result, the KM estimator is reconstructed as a special case of M-estimators. We 
derive the large-sample variance and the limiting distribution of the KM estimator 
in the spirit of M-estimation theory, allowing the establishment of
the corresponding confidence interval and confidence band. Simulation studies corroborate that the M-estimator under a quadratic M-function is exactly equivalent to the KM estimator and its asymptotic variance coincides with Greenwood's formula.
	
	The remainder of this paper is organized as follows. In Section 2, we define an M-estimator of 
a survival function and prove that the KM estimator matches with the 
M-estimator under a quadratic M-function. We derive the pointwise asymptotic variance and the joint 
limiting distribution of the KM estimator using M-estimation theory in Section 3. Various scenarios 
of simulations and real application in Section 4 demonstrate the equivalence relationship. 
Section 5 concludes with discussions.
	
\section{M-estimator of Survival Function}
	
\subsection{Problem Setup}\label{Setup}
	
We assume that the survival times to an event of interest are denoted by $T_1,\ldots,T_n$,
which are independently and identically distributed (i.i.d.) under a cumulative distribution function $F_0$ and
the corresponding survival function $S_0=1-F_0$. In a similar way, we assume i.i.d. 
censoring times $C_1,\ldots,C_n$ from a censoring distribution $G_0$.
The observed time of subject $i$ is $X_i=\min\{T_i,C_i\}$ with an indicator $\Delta_i=I\{T_i< C_i\}$ which equals 1 if the event of interest is observed before censoring and 0 otherwise. Often, independence is assumed between event time $T_i$ and censoring time $C_i$ for $i=1,\ldots,n$.
	Let $X_{(1)}<\cdots<X_{(K)}$ be the $K$ distinct observed event times. In what follows, we define the M-estimator of the
survival function and express the Kaplan--Meier estimator as a special case of the M-estimator.
	
	\subsection{M-estimator with Complete Data}
	
	We start with the case where there is no censoring (i.e., $\Delta_i=1$ for all $i$).  
Consider a known functional $m_{S}:\mathcal{S}\to \mathbb{R}$ where $\mathcal{S}=\{S(x):[0,\infty)\to [0,1];S(x)\text{ is nonincreasing}\}$. A popular method to find the estimator $\hat{S}(x)$
is to maximize a criterion function as follows,
	\begin{equation*}\label{M-estimator}
	\hat{S}(x)=\arg\max_{S(x)\in\mathcal{S}}M_n({S})=\arg\max_{S(x)\in\mathcal{S}}\frac{1}{n}\sum_{i=1}^{n}m_{S}(X_i).
	\end{equation*}
One special case of the M-function is the $L_2$ functional norm (or a quadratic norm) such that \begin{equation*}\label{m-function1}
m_{S}(X)=\int_{0}^\infty\big[-I\{X>x\}^2+2S(x)I\{X>x\}-S(x)^2\big]d\mu(x),
\end{equation*}
where $\mu(x)$ is a cumulative probability function.
	
	Let $\#\{i:\text{ Condition }\}$ be the number of observations that meet the condition. It is clear that when
the $L_2$ functional norm is  used, the empirical M-function is
	 \begin{equation}\label{M-function1}\begin{aligned} 
M_n(S)
&=\frac{1}{n}\sum_{i=1}^{n}\int_{0}^\infty\big[-I\{X_i>x\}^2+2S(x)I\{X_i>x\}-S(x)^2\big]
d\mu(x)\\
&=\int_{0}^\infty\Big[-\frac{\#\{i:X_i>x\}}{n}+2S(x)\frac{\#\{i:X_i>x\}}{n}-S(x)^2\Big]d\mu(x),\\
	\end{aligned}\end{equation}
	and the Kaplan--Meier estimator 
$$\begin{aligned}
	\hat{S}(x)&=\prod_{k:X_{(k)}\leq x}\Bigg(1-\frac{\#\{i:X_i= X_{(k)};\Delta_i=1\}}{\#\{i:X_i\geq X_{(k)}\}}\Bigg)\\&=\prod_{k:X_{(k)}\leq x}\Bigg(1-\frac{\#\{i:X_i= X_{(k)}\}}{\#\{i:X_i\geq X_{(k)}\}}\Bigg)\\&=\prod_{k:X_{(k)}\leq x}\Bigg(\frac{\#\{i:X_i>  X_{(k)}\}}{\#\{i:X_i\geq X_{(k)}\}}\Bigg)\\&=\prod_{k:X_{(k)}\leq x}\Bigg(\frac{\#\{i:X_i\geq X_{(k+1)}\}}{\#\{i:X_i\geq X_{(k)}\}}\Bigg)\\&=\frac{\#\{i:X_i> x\}}{n}
	\end{aligned}$$
is the maximizer of $M_n(S)$ in (\ref{M-function1}).
	
	\subsection{M-estimator with Censored Data}
	
When there are censored observations in the data, the empirical M-function of the observed data is
	\begin{equation}\label{M-function2}
	\begin{aligned}
	\widetilde{M}_n(S)&=\frac{1}{n}\sum_{i=1}^{n}\widetilde{m}_{S}(X_i,\Delta_i),\\
	\end{aligned}
	\end{equation}
	where
	\begin{equation*}\label{m-function2}
	\widetilde{m}_{S}(X,\Delta)=\begin{cases}
	m_{S}(X),&\Delta=1,\\
	\displaystyle\int_{0}^X\big[-I\{X>x\}^2+2S(x)I\{X>x\}-S(x)^2\big] d\mu(x),&\Delta=0.\\
	\end{cases}\end{equation*} 
	
	To obtain the optimizer, 
\begin{equation}\label{M-estimator2}
	\hat{S}(x)=\arg\max_{S(x)\in\mathcal{S}}\widetilde{M}_n({S}),
	\end{equation} 
we can apply the EM algorithm as follows:
\begin{itemize}
		\item \textbf{E-step}: Given the $g$th step estimator $\hat{S}^{(g)}(x)$, compute the expectation of the empirical M-function, 
\begin{equation}\label{E-step}
		\begin{aligned}E[M_n(S)|\hat{S}^{(g)}]=& \frac{1}{n}\sum_{\Delta_i=1}\int_{0}^\infty\big[-I\{X_i>x\}^2+2S(x)I\{X_i>x\}-S(x)^2\big]d\mu(x)\\
&+\frac{1}{n}\sum_{\Delta_i=0}\int_{0}^\infty\big[-\frac{\hat{S}^{(g)}(\max\{x, X_i\})}{\hat{S}^{(g)}(X_i)}+2S(x)\frac{\hat{S}^{(g)}(\max\{x, X_i\})}{\hat{S}^{(g)}(X_i)}-S(x)^2\big]d\mu(x).\\
		\end{aligned}
		\end{equation}
		\item \textbf{M-step}: Compute \begin{equation}\label{Update}
		\hat{S}^{(g+1)}(x)=\arg\max_{S(x)\in\mathcal{S}}E[M_n(S)|\hat{S}^{(g)}(x)].
		\end{equation}
	\end{itemize}
	The validity of this EM algorithm is guaranteed by Theorem \ref{EM}.
	\begin{thm}
		\label{EM}
		For all $\hat{S}^{(g)}(x)\in\mathcal{S}$, the quantity 
$E[M_n(S)|\hat{S}^{(g)}]-\widetilde{M}_n({S})$ is maximized when $S=\hat{S}^{(g)}$.
	\end{thm}
	
Based on Theorem \ref{EM}, we conclude that
$$\begin{aligned} \widetilde{M}_n({\hat{S}^{(g+1)}})&=E[M_n({\hat{S}^{(g+1)}})|\hat{S}^{(g)}]-\Big(E[M_n(\hat{S}^{(g+1)})|\hat{S}^{(g)}]-\widetilde{M}_n({\hat{S}^{(g+1)}})\Big)\\
	&\geq E[M_n({\hat{S}^{(g)}})|\hat{S}^{(g)}]-\Big(E[M_n(\hat{S}^{(g+1)})|\hat{S}^{(g)}]-\widetilde{M}_n({\hat{S}^{(g+1)}})\Big)\\
	&\geq E[M_n({\hat{S}^{(g)}})|\hat{S}^{(g)}]-\Big(E[M_n(\hat{S}^{(g)})|\hat{S}^{(g)}]-\widetilde{M}_n({\hat{S}^{(g+1)}})\Big)\\&=\widetilde{M}_n({\hat{S}^{(g)}})
	\end{aligned}$$
and thus the M-estimator would be obtained as the convergent point of the EM algorithm.
Through a proof by induction, it can be shown that the 
EM algorithm converges to the KM estimator, implying that the KM estimator is an M-estimator (\ref{M-estimator2}).
	
	\begin{thm}\label{Conver}
		If $\hat{S}^{(0)}$ is a non-increasing right-continuous function with $\hat{S}^{(0)}(0)=1$ and $\hat{S}^{(0)}(x)=\hat{S}^{(0)}(X_{(K)})$ for all $x\geq X_{(K)}$, the sequence of functions $\{\hat{S}^{(g)}\}$ with the recursive relation (\ref{Update}) for $g=0,1,\ldots$ satisfies that the limit function
\begin{equation}\label{conclusion}
		\hat{S}(x):=\lim\limits_{g\to\infty}\hat{S}^{(g)}(x)=\prod_{k:X_{(k)}\leq x}\Bigg\{1-\frac{\#\{i:X_i= X_{(k)};\Delta_i=1\}}{\#\{i:X_i\geq X_{(k)}\}}\Bigg\}.
		\end{equation}
That is, the EM algorithm would converge to the KM estimator.
	\end{thm}
	
	The proofs of Theorems \ref{EM} and \ref{Conver} are provided in the Appendix.
	
	\section{Asymptotic Properties of KM Estimator}
	
	Given that the KM estimator is an M-estimator (\ref{M-estimator2}), we can 
deduce its asymptotic properties in the spirit of M-estimation theory, including the 
asymptotic distribution, pointiwse confidence intervals, and the consequently asymptotic process
with the simultaneous confidence band.
	
	\subsection{Asymptotic Distribution}
	
Define
	$$\kappa_x(X,\Delta)=\begin{cases}
	I\{X>x\},&\Delta=1,\\
	\displaystyle\frac{S_0(\max\{x,X\})}{S_0(X)},&\Delta=0,
	\end{cases}$$ 
and we have
	\begin{equation*}
	E[\widetilde{m}_{S}(X,\Delta)|S_0]=\int_{0}^\infty\big[-\kappa_x(X,\Delta)+2S(x)\kappa_x(X,\Delta)-S(x)^2\big]d\mu(x)
	\end{equation*}
	and
	\begin{equation*}\frac{\partial E[\widetilde{m}_{S}(X,\Delta)|S_0]}{\partial S(x)}=2\kappa_x(X,\Delta)-2S(x).\end{equation*}
	
	Given that $X=\min\{T,C\}$ and indicator $\Delta=I\{T< C\}$ where $T\sim F_0=1-S_0$ and $C\sim G_0$, it is straightforward that $
	E\big\{\kappa_x(X,\Delta)\big\}=S_0(x)$. 
As a result, 
\begin{equation*}\begin{aligned}
	E\Bigg\{\frac{\partial E[\widetilde{m}_{S}(X,\Delta)|S_0]}{\partial S(x)}\Bigg\}=2S_0(x)-2S(x),
	\end{aligned}\end{equation*} 
which equals 0 when $S(x)=S_0(x)$. By the law of large numbers,
	$$\frac{\partial E[\widetilde{M}_n(S)|S_0]}{\partial S(x)}=\frac{1}{n}\sum_{i=1}^n\frac{\partial E[\widetilde{m}_{S}(X_i,\Delta_i)|S_0]}{\partial S(x)}$$ 
converges to $2S_0(x)-2S(x)$ in probability and thus $\hat{S}(x)$ converges to $S_0(x)$ in probability.
	
	If there exists an $X_F$ where $F_0(X_F)<1$, 
we have that when $0<x_1\leq x_2<X_F$,
	\begin{equation*}
	\begin{aligned}
	E\big\{\kappa_{x_1}(X,\Delta)\kappa_{x_2}(X,\Delta)\big\}&=S_0(x_2)(1-G_0(x_1))+S_0(x_1)S_0(x_2)\int_{0}^{x_1}\frac{dG_0(u)}{S_0(u)}
	\end{aligned}
	\end{equation*}
	by the conditional expectation formula. As a result, when $S=S_0$,
	\begin{equation*}
	\begin{aligned}
	&E\Bigg\{\frac{\partial E[\widetilde{m}_{S}(X,\Delta)|S_0]}{\partial S(x_1)}\frac{\partial E[\widetilde{m}_{S}(X,\Delta)|S_0]}{\partial S(x_2)}\Bigg\}\\
=&4E\{\big(\kappa_{x_1}(X,\Delta)-S(x_1)\big)\big(\kappa_{x_2}(X,\Delta)-S(x_2)\big)\}\\
=&4E\{\kappa_{x_1}(X,\Delta)\kappa_{x_2}(X,\Delta)\}-4S(x_1)S(x_2)\\
=&4S_0(x_1)S_0(x_2)\Biggl\{\frac{1-G_0(x_1)}{S_0(x_1)}+\int_{0}^{x_1}\frac{dG_0(u)}{S_0(u)}-1\Biggr\}\\
=&4S_0(x_1)S_0(x_2)\int_{0}^{x_1}\frac{1}{S_0^2(1-G_0)}d(1-S_0),
	\end{aligned}
	\end{equation*}
	which leads to the joint asymptotic distribution as 
\begin{equation}\label{Point1}\sqrt{n}\begin{pmatrix}
	\hat{S}(x_1)-S_0(x_1)\\	\hat{S}(x_2)-S_0(x_2)\\
	\end{pmatrix}\mathop{\to}^{\mathcal{D}}N\Biggl\{\begin{pmatrix}
	0\\0
	\end{pmatrix},\begin{pmatrix}
	\displaystyle S_0(x_1)^2\int_{0}^{x_1}\frac{d(1-S_0)}{S_0^2(1-G_0)}&\displaystyle S_0(x_1)S_0(x_2)\int_{0}^{x_1}\frac{d(1-S_0)}{S_0^2(1-G_0)}\\\displaystyle S_0(x_1)S_0(x_2)\int_{0}^{x_1}\frac{d(1-S_0)}{S_0^2(1-G_0)}&\displaystyle S_0(x_2)^2\int_{0}^{x_2}\frac{d(1-S_0)}{S_0^2(1-G_0)}
	\end{pmatrix}\Biggr\}.\end{equation}
	
	Under the log-transformation, the pointwise asymptotic distribution of $\log\hat{S}(x)$ is 
\begin{equation*}
\sqrt{n}(\log\hat{S}(x)-\log S_0(x))\mathop{\to}^{\mathcal{D}}N\Big\{0,\int_{0}^{x}\frac{1}{S_0^2(1-G_0)}d(1-S_0)\Big\},
\end{equation*}
	where the variance can be estimated by 
\begin{equation*}
\sum_{k:X_{(k)}\leq x}\frac{n\Delta_{(k)}}{\big(\sum_{l=k}^Kn_{(l)}\big)\big(\sum_{l=k+1}^Kn_{(l)}\big)}.
\end{equation*}
	
	\subsection{Confidence Band under Variance-Stabilizing Transformation}
	
	Given the asymptotic distribution (\ref{Point1}), the asymptotic distribution of 
the process $\hat{Z}(x)=\sqrt{n}\{\hat{S}(x)-S_0(x)\}$ ($0<x<X_F$) is 
provided in Theorem \ref{thm2}.
	
	\begin{thm}\label{thm2}
		\citep{Breslow1974,W.J.1980} If $F_0$ and $G_0$ are continuous and there exist a $X_F$ where $F_0(X_F)<1$, the process $\hat{Z}(x)=\sqrt{n}\{\hat{S}(x)-S_0(x)\}$ ($0<x<X_F$) converges weakly to a zero-mean 
Gaussian process ${Z}(x)$ with covariance function \begin{equation*}
		\label{Cov} {\rm cov}\big\{{Z}(x_1),{Z}(x_2)\big\}=\int_{0}^{x_1}\frac{S_0(x_1)S_0(x_2)}{S_0(u)^2(1-G_0(u))}d(1-S_0(u)),
		\end{equation*}
		where $0<x_1\leq x_2<X_F$.
	\end{thm}
	
	To deduce the confidence band by Theorem \ref{thm2},  
let $A(x)=\int_{0}^{x}{S_0(u)^{-2}(1-G_0(u))^{-1}}d(1-S_0(u))$, $H(x)=A(x)/\{1+A(x)\}$. It is clear that 
\begin{equation*}
\lim\limits_{x\to\infty}A(x)\geq\lim\limits_{x\to\infty}\int_{0}^{x}S_0(u)^{-2}d(1-S_0(u))=\infty
\end{equation*} 
and $\lim\limits_{x\to\infty}H(x)=1$.
For convenience, let $H(x)=1$ for $x\geq X_F$. Thus, the covariance function of 
the zero-mean Gaussian process ${Z}(x)$ in Theorem \ref{thm2} can be rewritten as 
\begin{equation*}
\label{Cov2}{\rm cov}\big\{{Z}(x_1),{Z}(x_2)\big\}=S_0(x_1)S_0(x_2)\frac{H(x_1)}{1-{H}(x_1)}
=S_0(x_1)S_0(x_2)\frac{H(x_1)(1-{H}(x_2))}{(1-{H}(x_1))(1-{H}(x_2))}.
	\end{equation*}
	Under the log-transformation, the process $\hat{Z}^*(x)=\sqrt{n}\{\log\hat{S}(x)-\log S_0(x)\}$ 
converges weakly to a zero-mean Gaussian process 
\begin{equation*}
\frac{B^0(H(x))}{1-{H}(x)},\quad 0<x<X_F,
\end{equation*}
	where $B^0(\cdot)$ is a standard Brownian bridge process on $[0,1]$. With the constant $$c_\alpha(a,b)=\inf \Biggl\{c:\mathbb{P}\Bigg(\sup_{[a,b]}|B^0(\cdot)|\leq c\Bigg)\geq 1-\alpha\Biggr\},\quad 0<\alpha<1,0<a<b<1,$$
	the asymptotic $(1-\alpha)$ confidence band of the survival function in the interval $[x_1,x_2]\subset[0,X_F]$ is
	\begin{equation*}
	\label{Confidence bands}	
\Bigg[\hat{S}(x)\exp\Bigg(-\frac{c_\alpha(\hat{H}(x_1),\hat{H}(x_2))}{\sqrt{n}(1-\hat{H}(x))}\Bigg),
\hat{S}(x)\exp\Bigg(\frac{c_\alpha(\hat{H}(x_1),\hat{H}(x_2))}{\sqrt{n}(1-\hat{H}(x))}\Bigg)\Bigg], \quad x_1<x<x_2,
	\end{equation*}
	where \begin{equation*}\hat{H}(x)=1-\Biggl\{1+\sum_{k:X_{(k)}\leq x}\frac{n\Delta_{(k)}}{\big(\sum_{l=k}^Kn_{(l)}\big)\big(\sum_{l=k+1}^Kn_{(l)}\big)}\Biggr\}^{-1}.\end{equation*}
	
	\section{Simulations and Application}
	
	\subsection{Synthetic Data}
	We conduct several simulation studies to compare the Kaplan--Meier estimator obtained by optimizing 
the M-function (\ref{M-function2}) with the existing maximum likelihood approach \citep{kaplan1958nonparametric}. 
We refer to the Kaplan--Meier estimator obtained by the proposed method as ``M-KM" and 
the existing maximum likelihood approach as ``KM". Both the estimation and inference, including the coverage probability of
confidence intervals and confidence bands, are studied. We consider two examples with $100$ 
replications for each.
	
	\textbf{Example 1.} For each sample of Example 1, $n=200$ observations of survival data are generated,
where $F_0$ and $G_0$ are exponential distributions with rates 1/3 and 1/6 respectively.

	\textbf{Example 2.}
	For each sample of Example 2, $n=500$ observations of survival data are generated, 
where $F_0$ and $G_0$ are Weibull$(1,1)$ and Weibull$(1,2)$ respectively.
	
\begin{figure}
		\centering
		\includegraphics[width=9cm]{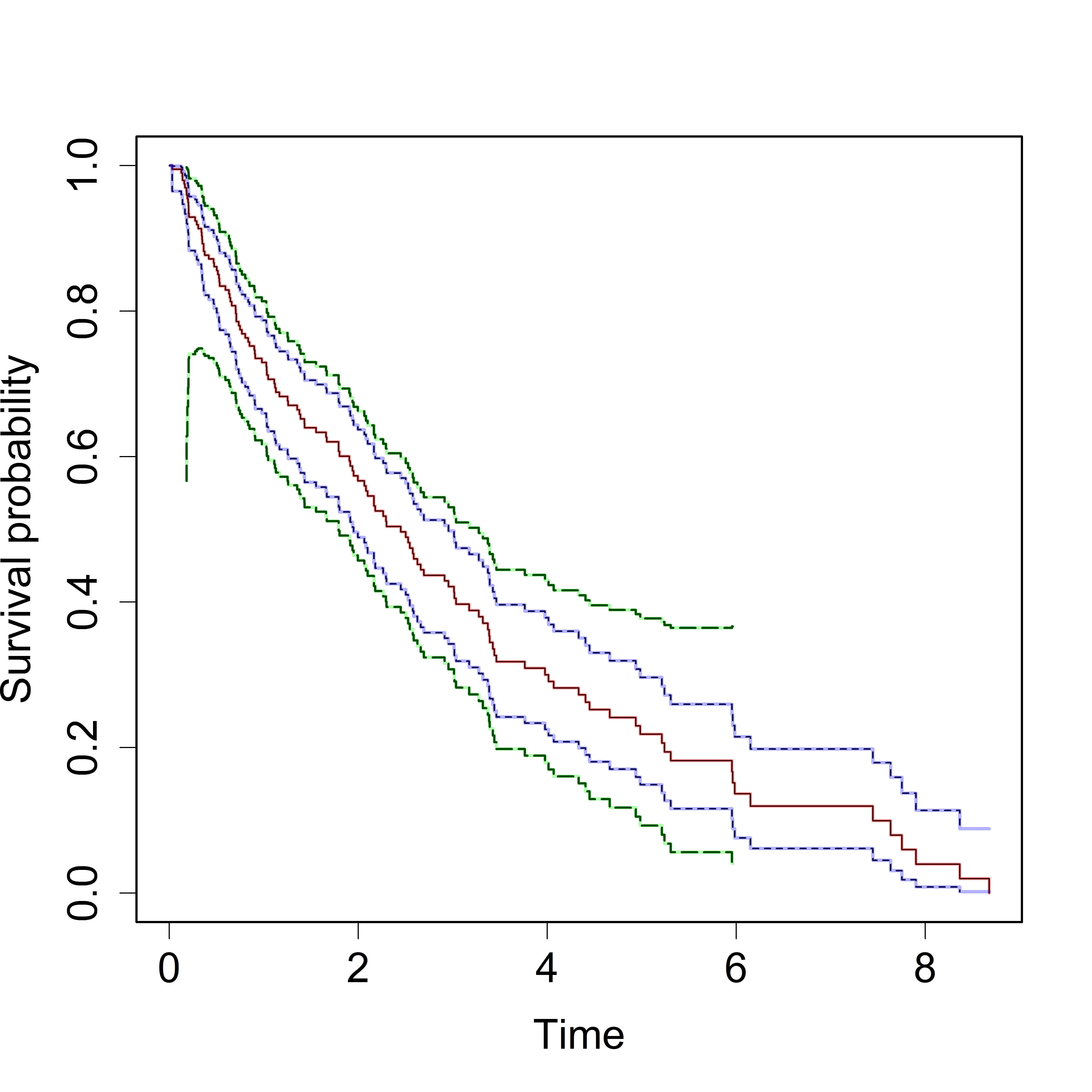}
		\includegraphics[width=9cm]{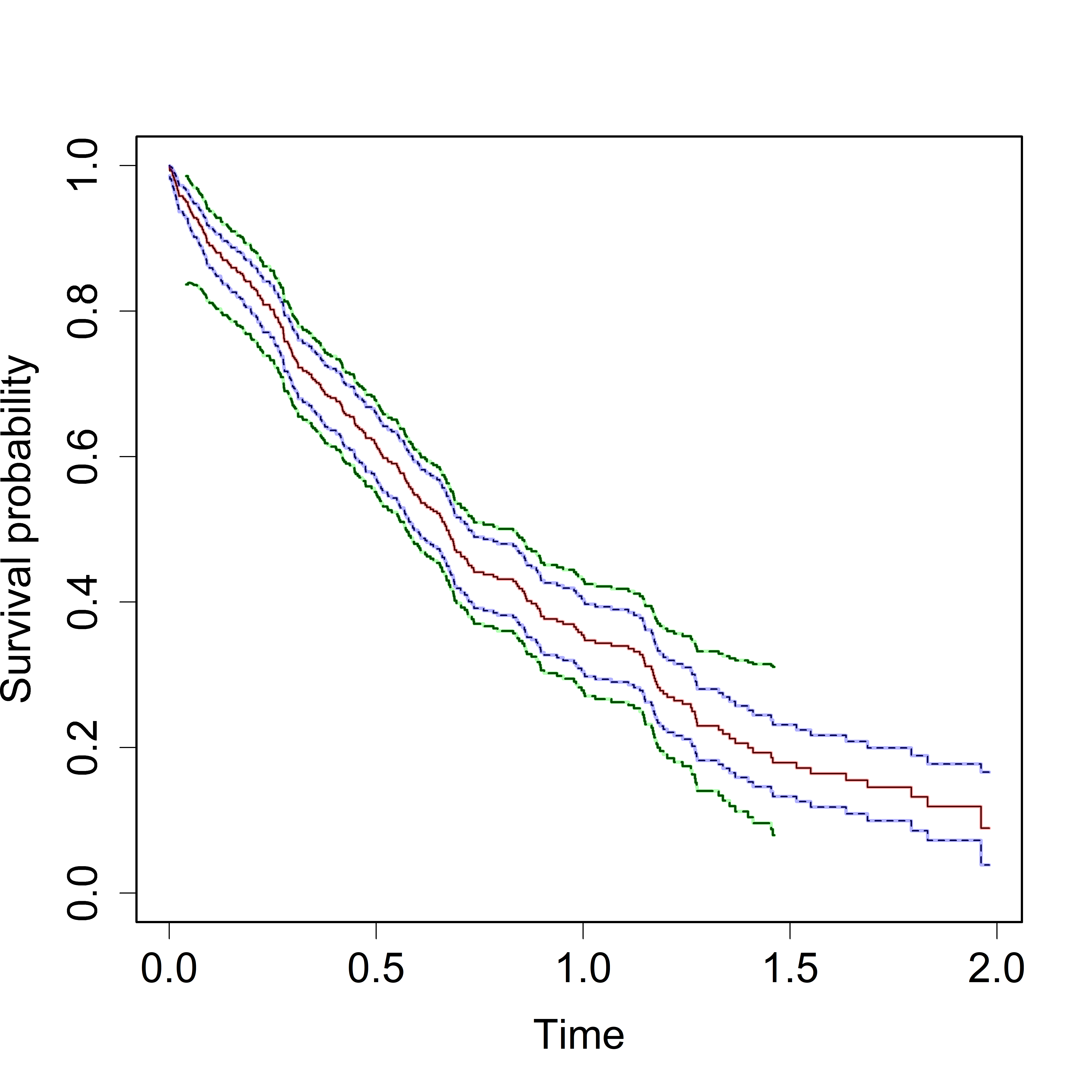}
		\caption{KM curves (red), 95\% confidence interval (blue) and 95\% confidence band (green) of Examples 1 (left) and 2 (right).}
		\label{fig-1}
	\end{figure}

	Figure \ref{fig-1} displays the KM estimators,
 pointwise 95\% confidence intervals and simultaneous 95\% confidence bands 
 computed by the proposed method and existing maximum likelihood approach 
 under one sample from Examples 1 and 2. It can be seen that although with different target functions, 
 the KM estimator by the proposed method is the same as the traditional one. The 95\% confidence intervals are 
 exactly the same as that by Greenwood's formula. In addition, the 95\% confidence band coincides 
 with the one proposed by \citet{W.J.1980}.
		
	Tables \ref{tab1} and \ref{tab2} show the coverage probability and the average length of the 95\% confidence interval  
at several time points over 100 repetitions for Examples 1 and 2. We choose seven time points $1,2,\ldots, 7$ in Example 1 and eight time points $0.1,0.3,\ldots,1.5$ in Example 2. Given that the KM estimator obtained by the proposed method equals the one computed via 
the maximum likelihood approach, the results of M-KM is the same as the existing one. The coverage probability of the 95\% confidence interval in Example 1 is around 0.95 at all seven time points. However, in Example 2, the coverage probability is  
less than 0.95 for larger time points. We also compute the coverage probability for the confidence bands. In Example 1, the coverage probability  of the confidence band is 0.97, while the value is 0.94 in Example 2. This means that the properties of the KM estimator can be successively recovered with the M-estimation theory, implying that KM estimator can be interpreted as an M-estimator.
	
		\begin{table}[htbp]
		\centering
		\caption{The coverage probability and the average length of the 95\% confidence interval at different time points over 100 
 replications under Example 1.}
		\begin{tabular}{lllllllll}
			\toprule
			&&\multicolumn{7}{c}{Time points}\\
			\cline{3-9}
			&   & 1     & 2     & 3     & 4     & 5     & 6     & 7 \\
			\midrule
			\multirow{2}[0]{*}{Coverage Probability}&M-KM    & 0.98  & 0.96  & 0.97  & 0.96  & 0.97  & 0.95  & 0.97 \\
			&KM       & 0.98  & 0.96  & 0.97  & 0.96  & 0.97  & 0.95  & 0.97 \\
			&&&&&&&&
			\\
			\multirow{2}[0]{*}{Length}&M-KM     & 0.1299 & 0.1516 & 0.1541 & 0.149 & 0.1401 & 0.1298 & 0.1196 \\
			&KM   & 0.1299 & 0.1516 & 0.1541 & 0.149 & 0.1401 & 0.1298 & 0.1196 \\
			\bottomrule
		\end{tabular}%
		\label{tab1}
	\end{table}%

		\begin{table}[htbp]
		\centering
		\caption{The coverage probability and the average length of the 95\% confidence interval at different time points over 100 
replications under Example 2.}
		\begin{tabular}{llllllllll}
			\toprule
			&&\multicolumn{8}{c}{Time points}\\
			\cline{3-10}
			&   & 0.1   & 0.3   & 0.5   & 0.7   & 0.9   & 1.1   & 1.3   & 1.5 \\
			\midrule
			\multirow{2}[0]{*}{Coverage Probability}&M-KM   & 0.94  & 0.92  & 0.88  & 0.88  & 0.85  & 0.89  & 0.91  & 0.84 \\
			&KM    & 0.94  & 0.92  & 0.88  & 0.88  & 0.85  & 0.89  & 0.91  & 0.84 \\
			&&&&&&&&&\\
			\multirow{2}[0]{*}{Length}&M-KM    & 0.052 & 0.079 & 0.0906 & 0.0962 & 0.0989 & 0.1005 & 0.1026 & 0.1078 \\
			&KM   & 0.052 & 0.079 & 0.0905 & 0.0962 & 0.0989 & 0.1005 & 0.1026 & 0.1078 \\
			\bottomrule
		\end{tabular}%
		\label{tab2}
	\end{table}%

	\subsection{Real Data Application}
	We apply the proposed method to two datasets to evaluate its performance. The first dataset is from 
a diabetic study, including $n=394$ observations and 8 variables.
	Among all observations, 197 patients are labeled ``high-risk" diabetic retinopathy according to 
the diabetic retinopathy study (DRS). For each patient, one eye was randomly chosen to receive 
the laser treatment and the other one received no treatment. The event of interest is that the visual acuity 
dropped below 5/200. Some records were censored due to death, dropout, or the end of the study. 
The second dataset is from a lung cancer study 
involving the North Central Cancer Treatment Group of 228 patients of advance lung cancer. 
The survival time was measured from initiation of the study to the time when the patient died or was censored.
	
\begin{figure}
		\centering
		\includegraphics[width=9cm]{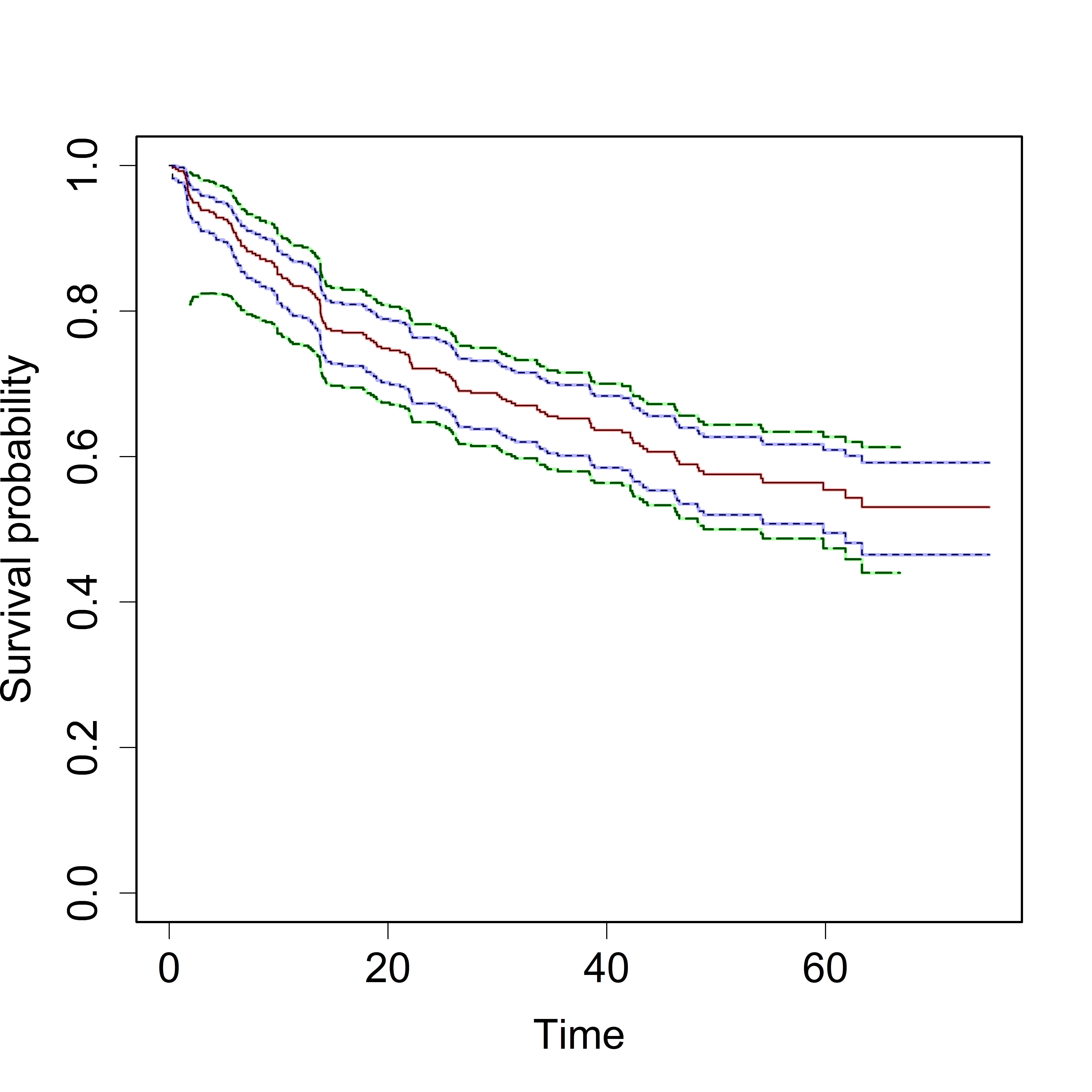}
		\includegraphics[width=9cm]{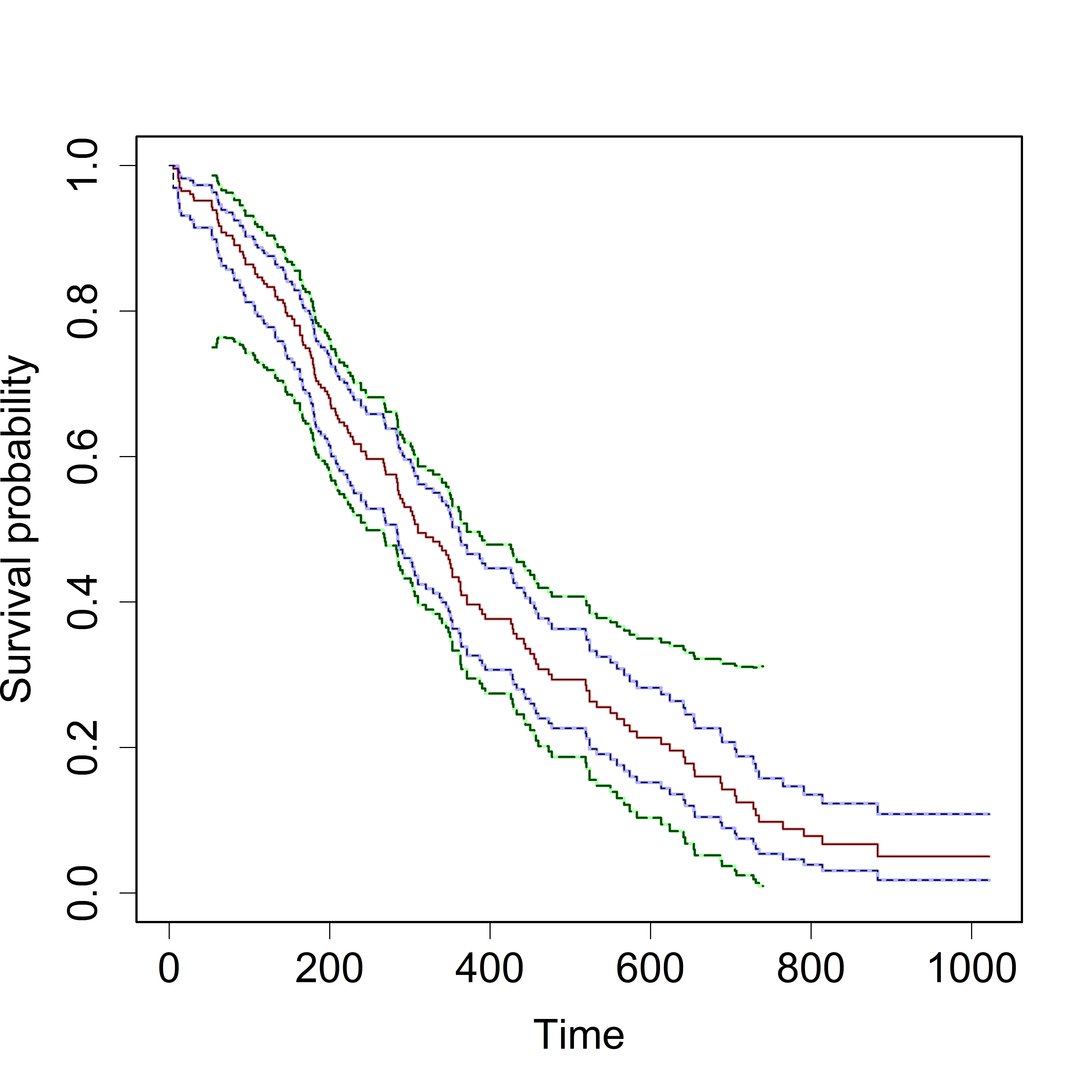}
		\caption{The KM curves (red), 95\% confidence interval (blue) and 95\% confidence band (green) 
for the diabetic (left) and lung cancer datasets (right).}
		\label{fig-2}
	\end{figure}

	The KM curves and the corresponding 95\% confidence intervals and 95\% confidence band computed by the proposed method and existing maximum likelihood approach for both datasets are shown in Figure \ref{fig-2}. Again, the estimation and inference result of the proposed method are the same as the existing ones, suggesting that the proposed method provides a 
new interpretation of the KM estimator as an M-estimator.
	
\section{Conclusions}

We reconstruct the KM estimator of the survival function as a special case of an M-estimator, which can be obtained recursively via 
the EM algorithm. The theoretical properties of the novel estimator, including the large-sample variance and the limiting distribution, are re-established in the spirit of M-estimation theory. Simulations and real application show that the reconstructed M-estimator is equivalent to the KM estimator under the quadratic M-function and both the consequent confidence interval and 
confidence band coincide with those obtained under Greenwood's formula.

Although we develop an M-estimator of the survival function, we only consider the quadratic M-function whose maximizer 
coincides with the KM estimator. It would be possible to develop other M-estimators of the survival function under different M-functions, such as the $L_{\rho}$-norm. In addition, since the proposed M-estimator is a nonparametric estimator for
the survival function, it is possible to develop its parametric counterpart, which may not be the same as the parametric maximum likelihood estimation. It is also interesting to investigate how such nonparametric M-estimator
can be utilized in fitting existing survival models with exogenous features, such as the Cox proportional hazard model.

\begin{appendices}
	\section{Proof of Theorem \ref{EM}}
		\rm
		By the definition of $E[M_n({\hat{S}^{(g)}})|\hat{S}^{(g)}]$, we can deduce that $$\begin{aligned}
		 &n\big(E[M_n(S)|\hat{S}^{(g)}]-\widetilde{M}_n({S})\big)\\=&\sum_{\Delta_i=0}\int_{0}^\infty\big[-\frac{\hat{S}^{(g)}(\max\{x, X_i\})}{\hat{S}^{(g)}(X_i)}+2S(x)\frac{\hat{S}^{(g)}(\max\{x, X_i\})}{\hat{S}^{(g)}(X_i)}-S(x)^2\big]d\mu(x)\\&-\int_{0}^{X_i}(-1+2S(x)-S(x)^2)d\mu(x)\\=&\sum_{\Delta_i=0}\int_{X_i}^\infty\big[-\frac{\hat{S}^{(g)}(x)}{\hat{S}^{(g)}(X_i)}+2S(x)\frac{\hat{S}^{(g)}(x)}{\hat{S}^{(g)}(X_i)}-S(x)^2\big]d\mu(x)\\=&\sum_{\Delta_i=0}\int_{X_i}^\infty-\big[S(x)-\frac{\hat{S}^{(g)}(x)}{\hat{S}^{(g)}(X_i)}\big]^2d\mu(x)+\sum_{\Delta_i=0}\int_{X_i}^\infty\frac{\hat{S}^{(g)}(x)^2}{\hat{S}^{(g)}(X_i)^2}d\mu(x)-\sum_{\Delta_i=0}\int_{X_i}^\infty\frac{\hat{S}^{(g)}(x)}{\hat{S}^{(g)}(X_i)}d\mu(x).\\
		\end{aligned}$$
Note that $n\big(E[M_n(S)|\hat{S}^{(g)}]-\widetilde{M}_n({S})\big)$ consists of three terms. The latter two terms are irrelevant to $S(x)$. By the Cauchy--Schwarz inequality, the former one is maximized if and only if $S(x)=\hat{S}^{(g)}(x)$.

	\section{Proof of Theorem \ref{Conver}}
		\rm Define $n_{(k)}=\#\{i:X_i= X_{(k)}\}$ and $\Delta_{(k)}=\#\{i:X_i= X_{(k)};\Delta_i=1\}$, we have that for $g=1,2,\ldots$, \begin{equation*}
		\begin{aligned}&E[M_n(S)|\hat{S}^{(g-1)}]\\=&
		 \frac{1}{n}\sum_{k=1}^{K}\int_{0}^\infty\Delta_{(k)}\big[-I\{X_{(k)}>x\}^2+2S(x)I\{X_{(k)}>x\}-S(x)^2\big]d\mu(x)\\
&+\frac{1}{n}\sum_{k=1}^{K}\int_{0}^\infty(n_{(k)}-\Delta_{(k)})
\Big[-\frac{\hat{S}^{(g-1)}(\max\{x, X_{(k)}\})}{\hat{S}^{(g-1)}(X_{(k)})}+2S(x)\frac{\hat{S}^{(g-1)}(\max\{x, X_{(k)}\})}{\hat{S}^{(g-1)}(X_{(k)})}-S(x)^2\Big]d\mu(x).\\
		\end{aligned}
		\end{equation*}
		Hence, 
\begin{equation*}\label{Update2}
		\begin{aligned}
		\hat{S}^{(g)}(x)=&\arg\max_{S(x)\in\mathcal{S}}E[M_n(S)|\hat{S}^{(g-1)}]\\
		=&\frac{1}{n}\sum_{k=1}^{K}\big[\Delta_{(k)}I\{X_{(k)}>x\}+(n_{(k)}-\Delta_{(k)})\frac{\hat{S}^{(g-1)}(\max\{x, X_{(k)}\})}{\hat{S}^{(g-1)}(X_{(k)})}\big].
		\end{aligned}
		\end{equation*}
		If $\hat{S}^{(g-1)}$ is a non-increasing right-continuous function with $\hat{S}^{(g-1)}(0)=1$, it is clear that $I\{X_{(k)}>x\}$ and $\hat{S}^{(g-1)}(\max\{x, X_{(k)}\})/\hat{S}^{(g-1)}(X_{(k)})$ are both non-increasing right-continuous functions and thus $\hat{S}^{(g)}$ is a non-increasing right-continuous function with $\hat{S}^{(g)}(0)=1$. Given $\hat{S}^{(0)}$ is a non-increasing right-continuous function with $\hat{S}^{(0)}(0)=1$, by induction, we have $\hat{S}(x)$ is a non-increasing right-continuous function with $\hat{S}(0)=1$.
		
		\begin{itemize}
			\item For all $x\in[0,X_{(1)})$, it is obvious that $\hat{S}^{(g)}(x)=1$ for $g=0,1,\ldots$ and equation (\ref{conclusion}) holds.
			\item We then prove that equation (\ref{conclusion}) holds at $x=X_{(1)},\ldots,X_{(K)}$.
			\begin{itemize}
				\item It is clear that for $g=0,1,\ldots$, $$\begin{aligned}
				 \hat{S}^{(g)}(X_{(1)})&=\frac{1}{n}\big[\sum_{k=1}^K(n_{(k)}-\Delta_{(k)})+\sum_{k=2}^K\Delta_{(k)}\big]\\&=1-\frac{\Delta_{(1)}}{\sum_{k=1}^Kn_{(k)}}\\&=1-\frac{\#\{i:X_i= X_{(1)};\Delta_i=1\}}{\#\{i:X_i\geq X_{(1)}\}}\\
				\end{aligned}$$ and equation (\ref{conclusion}) holds at $x=X_{(1)}$.
				\item  Suppose that equation (\ref{conclusion}) holds at $x=X_{(l-1)}$ ($l=2,\ldots,K$). If $\hat{S}(X_{(l-1)})=0$, $\hat{S}(X_{(l)})=0$ and thus equation (\ref{conclusion}) holds at $x=X_{(l)}$. If $\hat{S}(X_{(l-1)})\neq0$, by the convergence condition of the EM algorithm, 
\begin{equation}\label{Convergence}
				\begin{aligned}
				\hat{S}(x)=&\arg\max_{S(x)\in\mathcal{S}}E[M_n(S)|\hat{S}]\\
				=&\frac{1}{n}\sum_{k=1}^{K}\big[\Delta_{(k)}I\{X_{(k)}>x\}+(n_{(k)}-\Delta_{(k)})\frac{\hat{S}(\max\{x, X_{(k)}\})}{\hat{S}(X_{(k)})}\big],
				\end{aligned}
				\end{equation}
				we have \begin{equation*}
				\label{Ratio}\begin{aligned}
				\frac{\hat{S}(X_{(l)})}{\hat{S}(X_{(l-1)})}
				 =&\frac{n_{(l)}-\Delta_{(l)}+\sum_{k=l+1}^{K}n_{(k)}+\hat{S}(X_{(l)})\sum_{k=1}^{l-1}{(n_{(k)}-\Delta_{(k)})}/{\hat{S}(X_{(k)})}}{\sum_{k=l}^{K}n_{(k)}+\hat{S}(X_{(l-1)})\sum_{k=1}^{l-1}{(n_{(k)}-\Delta_{(k)})}/{\hat{S}(X_{(k)})}}.
				\end{aligned}
				\end{equation*}
				It is clear that $$	\begin{aligned}
				 \frac{\hat{S}(X_{(l)})}{\hat{S}(X_{(l-1)})}&=\frac{n_{(l)}-\Delta_{(l)}+\sum_{k=l+1}^{K}n_{(k)}}{\sum_{k=l}^{K}n_{(k)}}\\&=1-\frac{\Delta_{(l)}}{\sum_{k=l}^Kn_{(k)}}\\&=1-\frac{\#\{i:X_i= X_{(l)};\Delta_i=1\}}{\#\{i:X_i\geq X_{(l)}\}}\\
				\end{aligned}$$ and equation (\ref{conclusion}) holds at $x=X_{(l)}$.
			\end{itemize}
			\item Finally, we prove that equation (\ref{conclusion}) holds for $x\in(X_{(l-1)},X_{(l)})$ ($l=2,\ldots,K$) and for $x\in(X_{(K)},\infty)$.
			\begin{itemize}
				\item For $x\in(X_{(l-1)},X_{(l)})$ ($l=2,\ldots,K$), if $\hat{S}(X_{(l-1)})=0$, $\hat{S}(x)=0$ and thus equation (\ref{conclusion}) holds; otherwise, by the convergence condition (\ref{Convergence}) of the EM algorithm, we have \begin{equation*}
				\label{Ratio2}\begin{aligned}
				\frac{\hat{S}(x)}{\hat{S}(X_{(l-1)})}
				 =&\frac{n_{(l)}-\Delta_{(l)}+\sum_{k=l+1}^{K}n_{(k)}+\hat{S}(x)\sum_{k=1}^{l-1}{(n_{(k)}-\Delta_{(k)})}/{\hat{S}(X_{(k)})}}{\sum_{k=l}^{K}n_{(k)}+\hat{S}(X_{(l-1)})\sum_{k=1}^{l-1}{(n_{(k)}-\Delta_{(k)})}/{\hat{S}(X_{(k)})}},
				\end{aligned}
				\end{equation*}
				implying that $\hat{S}(x)/\hat{S}(X_{(l-1)})=1$. Thus, equation (\ref{conclusion}) holds.
				\item For $x\in(X_{(K)},\infty)$, if $\hat{S}(X_{(K)})=0$, $\hat{S}(x)=0$ and thus equation (\ref{conclusion}) holds; otherwise, it is clear that for $g=1,2,\ldots$, \begin{equation*}
				\label{Ratio3}\begin{aligned}
				\frac{\hat{S}^{(g)}(x)}{\hat{S}^{(g)}(X_{(K)})}
				 =&\frac{\sum_{k=1}^{K}(n_{(k)}-\Delta_{(k)})\hat{S}^{(g-1)}(x)}{\sum_{k=1}^{K}(n_{(k)}-\Delta_{(k)})\hat{S}^{(g-1)}(X_{(K)})}\\
=&\frac{\hat{S}^{(g-1)}(x)}{\hat{S}^{(g-1)}(X_{(K)})}\\=&\cdots\\=&\frac{\hat{S}^{(0)}(x)}{\hat{S}^{(0)}(X_{(K)})}\\=&1.
				\end{aligned}
				\end{equation*}
				Therefore, $$\frac{\hat{S}(x)}{\hat{S}(X_{(K)})}=\lim\limits_{g\to\infty}\frac{\hat{S}^{(g)}(x)}{\hat{S}^{(g)}(X_{(K)})}=1,$$ and equation (\ref{conclusion}) holds.
			\end{itemize}
		\end{itemize}
\end{appendices}

\bibliography{thesis}
\bibliographystyle{apalike}
\end{document}